\documentclass[journal=jpclcd,manuscript=letter]{achemso}
\usepackage{graphicx}
\usepackage{epstopdf}
\usepackage{tabularx}
\usepackage{graphicx}

\title{Caloric Effects in Methylammonium Lead Iodide from Molecular Dynamics Simulations }
\author{Shi Liu}
\email{sliu@carnegiescience.edu}
\affiliation{Extreme Materials Initiative, Geophysical Laboratory, Carnegie Institution for Science, Washington, D.C. 20015-1305 USA}
\author{Ronald E. Cohen}
\email{rcohen@carnegiescience.edu}
\affiliation{Extreme Materials Initiative, Geophysical Laboratory, Carnegie Institution for Science, Washington, D.C. 20015-1305 USA}
\alsoaffiliation{Department of Earth- and Environmental Sciences, Ludwig Maximilians Universit\"{a}t, Munich 80333, Germany}

\begin{document}
\abstract{
Organic-inorganic hybrid perovskite architecture could serve as a robust platform for materials design to realize functionalities beyond photovoltaic applications. We explore caloric effects in organometal halide perovskites, taking methylammonium lead iodide (MAPbI$_3$) as an example, using all-atom molecular dynamics simulations with a first-principles based interatomic potential. The adiabatic thermal change is estimated directly by introducing different driving fields in the simulations. We find that MAPbI$_3$ exhibits both electrocaloric and mechanocaloric effects at room temperature. Local structural analysis reveals that the rearrangement of molecular cations in response to electric and stress fields is responsible for the caloric effects. The enhancement of caloric response could be realized through strain engineering and chemical doping. }
\newpage
\maketitle

\section{Introduction}
Organometal halide perovskites (OMHPs) are promising candidates for low-cost-to-power photovoltaic technologies. They are now at the frontier of renewable energy research for their record speed of increasing power conversion efficiency which exceeded 20\% after just 6 years of research~\cite{Kojima09p6050,Park13p2423,Stranks15p391,KRICT}. Since OMHP crystals are easily grown~\cite{Saidamino2015}, they could serve as a robust platform as active materials beyond photovoltaic applications. Represented by methylammonium lead iodide (MAPbI$_3$), these materials have the $ABX_3$ perovskite structure with an organic monovalent cation at $A$-site, a divalent metal at $B$-site, and a halide anion at $X$-site. The molecular cation has a permanent dipole and interacts with the inorganic $BX_3$ scaffold mostly through van der Waals interactions and hydrogen bonding~\cite{Wang14p1424,Egger14p2728}. Compared to inorganic perovkites, one unique structural feature of OMHPs comes from the $A$-site organic molecules that may undergo significant rearrangement at room temperature due to the small rotational barrier (as low as several meV)~\cite{Brivio13p042111,Motta15p7026}. Alignment of molecular dipoles could give rise to spontaneous polarization and polar domains, which may modulate the band gap and suppress electron-hole recombination, essential for the superior photovoltaic performance of OMHPs~\cite{Frost14p2584,Liu15p693,Ma15p248}. Though switchable polar domains in $\beta$-MAPbI$_3$ have been observed in some studies of piezoresponse force microscopy (PFM)~\cite{Kutes14p3335}, the robustness of the room-temperature ferroelectricity in MAPbI$_3$ is still under debate~\cite{Xiao15p193,Fan15p1155,Kim15p1729,Coll15p1408}. Further functionality of OMHPs has been proposed, ranging from thermoelectrics,~\cite{He14p5394,Mettan15p11506} spintronics~\cite{Kim14p6900,Zheng15p7794} to light-emitting diodes and lasers.~\cite{Zhang14p5995,Zhu15p636}

There has been a surge of interest in using caloric materials to develop environmentally friendly solid-state cooling technology as an alternative of traditional vapour-compression technology that relies on high global-warming potential refrigerants (hydrofluorocarbons and hydrochlorofluorocarbons)~\cite{Scott11p229,Takeuchi15p48,Moya14p439,Valant12p980,Rose12p187604}. Caloric effects refers to the phenomena in which the temperature of the materials changes in response to the change of external driving field~\cite{Moya14p439}. There are different types of caloric effects depending on the nature of the driving field. The electrocaloric effect (ECE) is driven by an applied electric field, whereas the mechanocaloric effect is induced by applied mechanical stress (the elastocaloric effect is driven by a uniaxial stress and the barocaloric effect is driven by an isotropic stress). Magnetocaloric materials will have reversible thermal changes in response to changes of applied magnetic field.~\cite{Moya14p439} A cooling cycle is realized through two constant-entropy (adiabatic) transitions and two constant driving-force transitions (Figure 1)~\cite{Scott11p229}. It is known that the ECE could occur in any insulator with a large, temperature dependent, dielectric susceptibility.~\cite{Rose12p187604} The intrinsic dipoles afforded by the molecular cations may give rise to ECE in OMHPs. In this work, we explore various caloric effects of MAPbI$_3$ with full-atom molecular dynamics (MD) simulations. We study both electrocaloric and mechanocaloric effects under different temperatures. The temperature change is estimated directly by adiabatically applying driving fields in MD simulations. We find that MAPbI$_3$ has weak though non-zero ECE under high electric fields ($E>1$~MV/cm), and applying epitaxial strain could enhance the ECE. Interestingly, our simulations point out that MAPbI$_3$ has decent temperature-insensitive mechanocaloric response comparable to some shape-memory alloys of Ni-Ti.

\section{Computational Methods}
We carried out MD simulations on a $20\times20\times20\times20$ perovskite-type supercell (96000 atoms) using an interatomic model potential derived from first-principles~\cite{Mattoni15p17421}. The model potential of MAPbI$_3$ consists of organic-organic, organic-inorganic, and inorganic-inorganic interactions. The organic-organic interaction includes both the intramolecular and intermolecular interactions described by the standard AMBER force field. The interactions within the inorganic Pb--I scaffold are described by Buckingham potential. The interactions between the MA cation and the PbI$_3$ sublattice are described as the sum of Buckingham, electrostatic, and Lennard-Jones potentials (see details in ref.\citenum{Mattoni15p17421}). The temperature is controlled by a Nos{\'e}-Hoover thermostat and the pressure is maintained at 0.0 MPa by the Parrinello-Rahman barostat implemented in LAMMPS~\cite{Plimpton95p1}. A 0.5~fs timestep is used and the system is equilibrated for more than 2~ns before turning on the external driving field. During the application of the driving field, a large inertial parameter $M_s$ is used for the thermostat (Tdamp = 5 ns in LAMMPS) to prevent heat transfer between the system and the thermostat, simulating an adiabatic process. 

\section{Results and Discussions}
Figure 2(a) shows the adiabatic thermal changes upon the application of electric fields of different magnitudes along the $x$ direction. The electric field is turned on gradually over a 10~ps period. We find that even with a high electric field (1~MV/cm), the induced change in temperature ($\Delta T$) is relatively small ($\approx$0.27~K), compared to $\Delta T= 12~K$ observed in PbZr$_{0.95}$Ti$_{0.05}$O$_3$ thin films on application of electric fields of 0.48~MV/cm~\cite{Mischenko06p1270}. Increasing the magnitude of electric field to 4~MV/cm results in a larger $\Delta T$ of 4.1~K. Though higher fields up to 1.4~GV/m have been achieved in ultrathin crystal of  barium titanate~\cite{Morrison05p152903}, it is likely to be a challenge for MAPbI$_3$ that has a small band gap to withstand ultrahigh electric fields. We further investigate the temperature dependence of $\Delta T$ from 250 to 350~K. As shown in Figure 2(b), the $\Delta T$ remains almost constant for a given applied field. It is known that for normal ferroelectrics, $\Delta T$ in general will peak around the phase transition temperature ($T_c$) and will shift toward a higher temperature as the electric field increases~\cite{Rose12p187604}. The weak temperature dependence of $\Delta T$ is usually found in relaxor ferroelectrics~\cite{Goupil14p1301688}. This hints the structural similarity between MAPbI$_3$ and relaxor ferroelectrics with polar nanoregions.  

To provide an atomistic description of the ECE in MAPbI$_3$, we analyze the distribution of molecular dipoles along the Cartesian axes (Figure 3).  In the absence of an electric field, the distribution is symmetric and broad for all three components of molecular dipoles ($\mu_x$, $\mu_y$ and $\mu_z$), indicating an isotropic distribution of molecular orientations (Figure 3 inset, $E$ = 0~MV/cm). By applying an electric field along the $+x$ direction, the distribution of $\mu_x$ is heavily skewed in that direction. This corresponds to the formation of a more ordered structure with dipoles aligning with the external electric field (Figure 3 inset, $E$ = 4~MV/cm) . Due to the conservation of total entropy in an adiabatic reversible process, the decrease of the configurational entropy (that characterizes the degree of order of atomic positions) is compensated by the increase of thermal entropy (that characterizes lattice vibrations), thus giving rise to the increase of temperature. 

Previous studies have demonstrated that the epitaxial misfit strain can significantly affect the phase boundaries in ferroelectric thin films~\cite{Pertsev98p1988,Choi04p1005,Alpay04p8118}, which could be used to tune the electrocaloric response~\cite{Akcay08p024104,Liu15p032901}. We then explore the effect of epitaxial strain on the ECE of MAPbI$_3$. In our MD simulations, a 2\% and 4\% compressive strain ($\eta$) is applied in $yz$ plane, respectively, with the electric field applied along the $x$ axis (the $x$-dimension of the supercell is allowed to relax). The results are presented in Figure 4a. We find that the compressive epitaxial strain enhances the ECE by at least 2--3 times. For instance, $\Delta T(\eta = 4\%)$ for $E=2$~MV/cm at 300~K is about 5.2~K, which is about 4 times larger than that (1.3 K) in the free-standing case. Similarly, with a higher field (4 MV/cm), the temperature change can go up to 8.6~K under a 4\% compressive epitaxial strain. Figure 4b shows the effect of epitaxial strain on the probability distribution of $\mu_x$. We find that in the strained MAPbI$_3$ ($\eta = 4$\%, $E$ = 0 MV/cm), there are more molecular dipoles aligning along the $x$ axis, as revealed by the enhanced peaks at both $\mu_x = -1.7$ and $1.7$ D and the suppressed peak at $\mu_x = 0$~D. Furthermore, compared to the freestanding MAPbI$_3$ under the same electric field, the epitaxially-strained MAPbI$_3$ is closer to a single domain, indicated by the nearly-zero probability of $\mu_x  < 0$~D. Therefore, the change of local polarization is significantly larger in strained MAPbI$_3$, responsible for the larger adiabatic thermal change. 

It is noted that the chemical bonds (Pb--I) in MAPbI$_3$ are not stiff~\cite{Egger16ASAP}, because of the large lattice constants (5.7--6.3~\AA, vs 3.8--4.2 \AA~common for oxide perovskites) and low oxidation state of halide anions (-1, vs -2 in oxide perovskites), giving rise to a bulk moduli (10-25 GPa) that is much smaller (at least tens of GPa) than values of typical oxide perovskites~\cite{Feng14p081801,Sun15p18450}. This suggests that MAPbI$_3$ is a ``soft" solid and could be stretched/compressed with even a moderate stress. We examine the mechanocaloric response of MAPbI$_3$ to both uniaxial ($\Delta \sigma_u$) and isotropic stress ($\Delta \sigma_i$). In MD simulations, the uniaxial stress is applied by slowly increasing the cell dimension along the $x$ axis while the dimensions along $y$ and $z$  directions are free to relax with stress maintained at 0 GPa. Figure 5a presents the temperature profiles in response to uniaxial stress of different loading speed and final magnitudes. It is found that stretching MAPbI$_3$ by the same amount with two different speeds (blue and orange lines in Figure 5a) results in similar changes in temperature ($\approx 3.5$~K for $\Delta \sigma_u=0.45$~GPa). Most notably, $\Delta T$ for $\Delta \sigma_u=0.55$~GPa is about 10.7~K, which is almost comparable to shape-memory alloys of Ni-Ti~\cite{Moya14p439,Cui12p101}. The evolution of the probability distribution functions of $\mu_x$, $\mu_y$ and $\mu_z$ under $\Delta \sigma_u=0.45$~GPa is shown in Figure 5b. We find that the rotation and alignment of molecular dipoles along the $x$ axis, as characterized by the decreasing $P(\mu_x = 0~{\rm D} )$ and increasing $P(\mu_x = \pm 1.7~{\rm D} )$, is responsible for the temperature change. We also evaluate the values of $\Delta T$ for $\Delta \sigma_u=0.45$~GPa at temperatures 250, 280, 300, 320, and 350~K, and find that the thermal changes vary from 3.3--3.8~K, showing weak temperature dependence. The heating caused by the isotropic compression is also significant (Figure 5c) that the temperature increases by 4.5~K for $\Delta \sigma_i=0.57$~GPa, which corresponds to a 1\% change in lattice constants. Under an isotropic stress, the probability distribution functions of molecular dipoles (Figure 5d) remain almost unchanged. This suggests that the increase in temperature is mainly due to the reduction of configuration entropy of a smaller volume. 
 
Finally, we show the cooling steps upon the removal of various driving forces in Figure 6. Before removing the driving force, the supercell is first equilibrated to 300~K (heat ejection step). When the electric/stress field is gradually removed, the temperature goes down, eventually reaching a state with lower temperature, which could be used to load heat for cooling. Though previous studies show that MAPbI$_3$ has large dielectric loss~\cite{Lin14p106} which may lead to heating during charging/discharging, recent experiments demonstrate that crystalized MAPbI$_3$ can be optimized to small dielectric loss of 0.02 at 1MHz~\cite{Kim16p756}. 

 \section{Conclution}
In summary, our molecular dynamics simulations demonstrate that MAPbI$_3$ exhibits many typical caloric effects, which is due to the rearrangement of $A$-site molecular cations in response to electric and stress fields. Given the soft chemical bonds in MAPbI$_3$, we expect the mechanocaloric effect is more likely to be observed experimentally. Organometal halide perovskites have demonstrated structural flexibility with varying atomic compositions and phase dimensionality~\cite{Zheng15p4862}. The caloric response in these materials could be further enhanced through chemical doping. For example, replacing the CH$_3$NH$_3^+$ with a more polar cation is likely to increase the electrocaloric effect.\\

\noindent {\bf Acknowledgements} This work is partly supported by US ONR. 
SL and REC are supported by the Carnegie Institution for Science. REC is also support by the
ERC Advanced Grant ToMCaT. Computational support was provided by the US DOD through a Challenge Grant from the HPCMO.

\newpage
\
\providecommand{\latin}[1]{#1}
\providecommand*\mcitethebibliography{\thebibliography}
\csname @ifundefined\endcsname{endmcitethebibliography}
  {\let\endmcitethebibliography\endthebibliography}{}

\newpage
\begin{figure}
\centering
\includegraphics[scale=1.0]{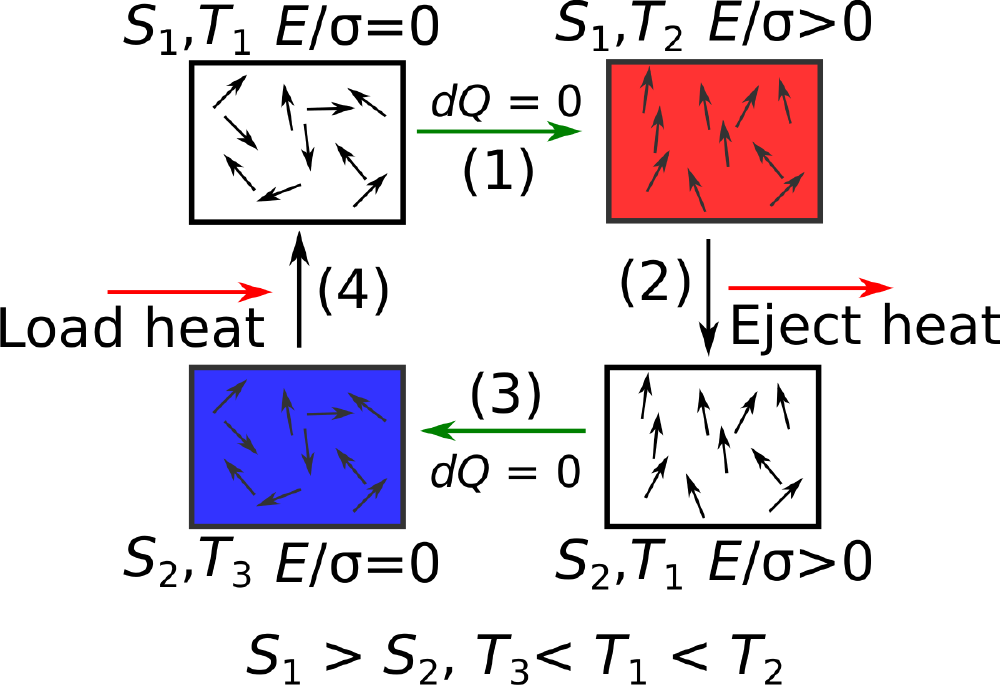}\\
 \caption{Schematic of a caloric cooling cycle involving two adiabatic processes (steps 1 and 3) and two constant driving-field ($E$: electric field, $\sigma$: stress field) processes (steps 2 and 4).}
 \end{figure}
 
\begin{figure}
\centering
\includegraphics[scale=1.0]{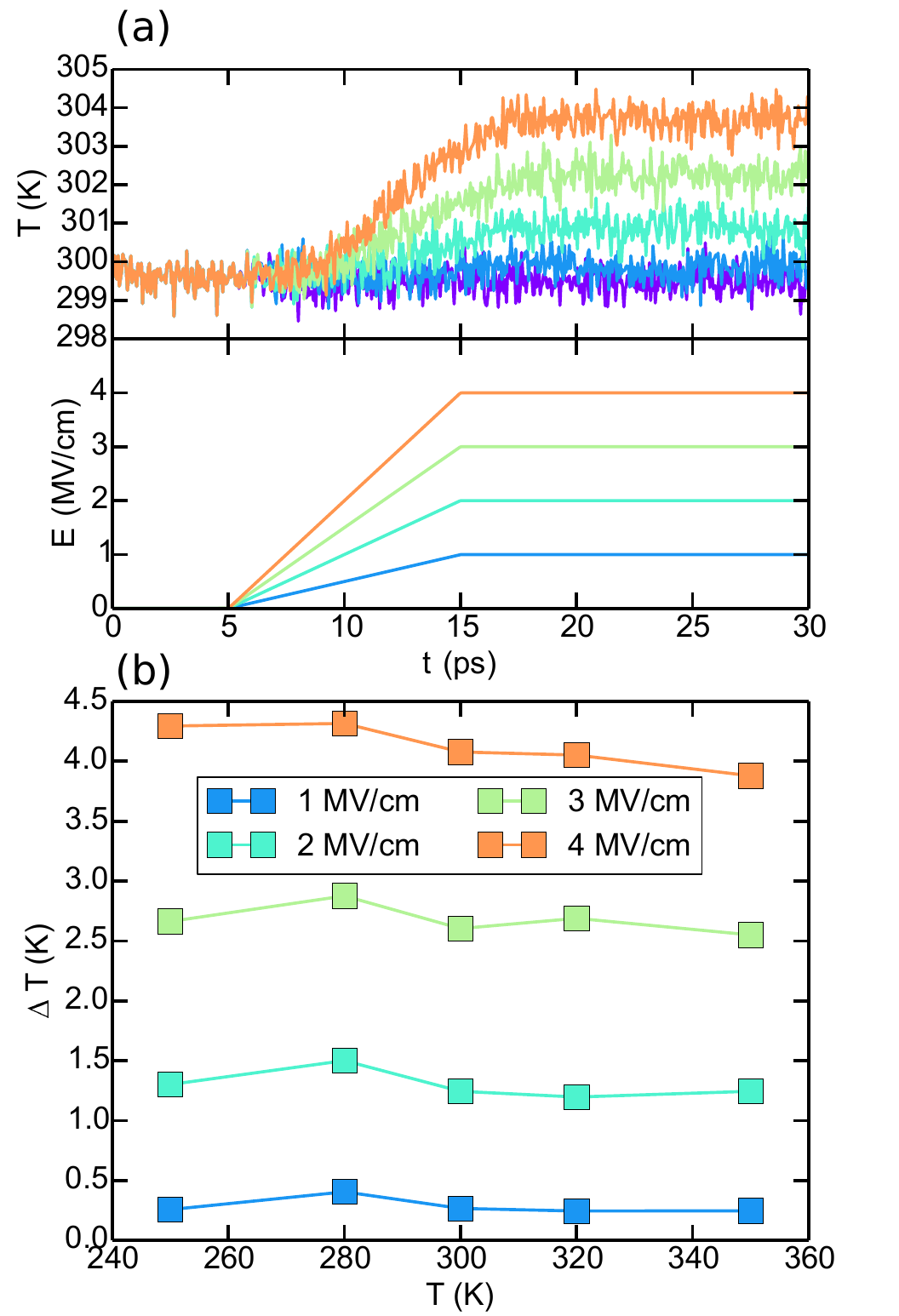}\\
 \caption{Electrocaloric effect in MAPbI$_3$. (a) Adiabatic thermal change in response to electric fields of different magnitudes. (b) Temperature and field dependence of electrocaloric effect. }
 \end{figure}

\begin{figure}
\centering
\includegraphics[scale=0.8]{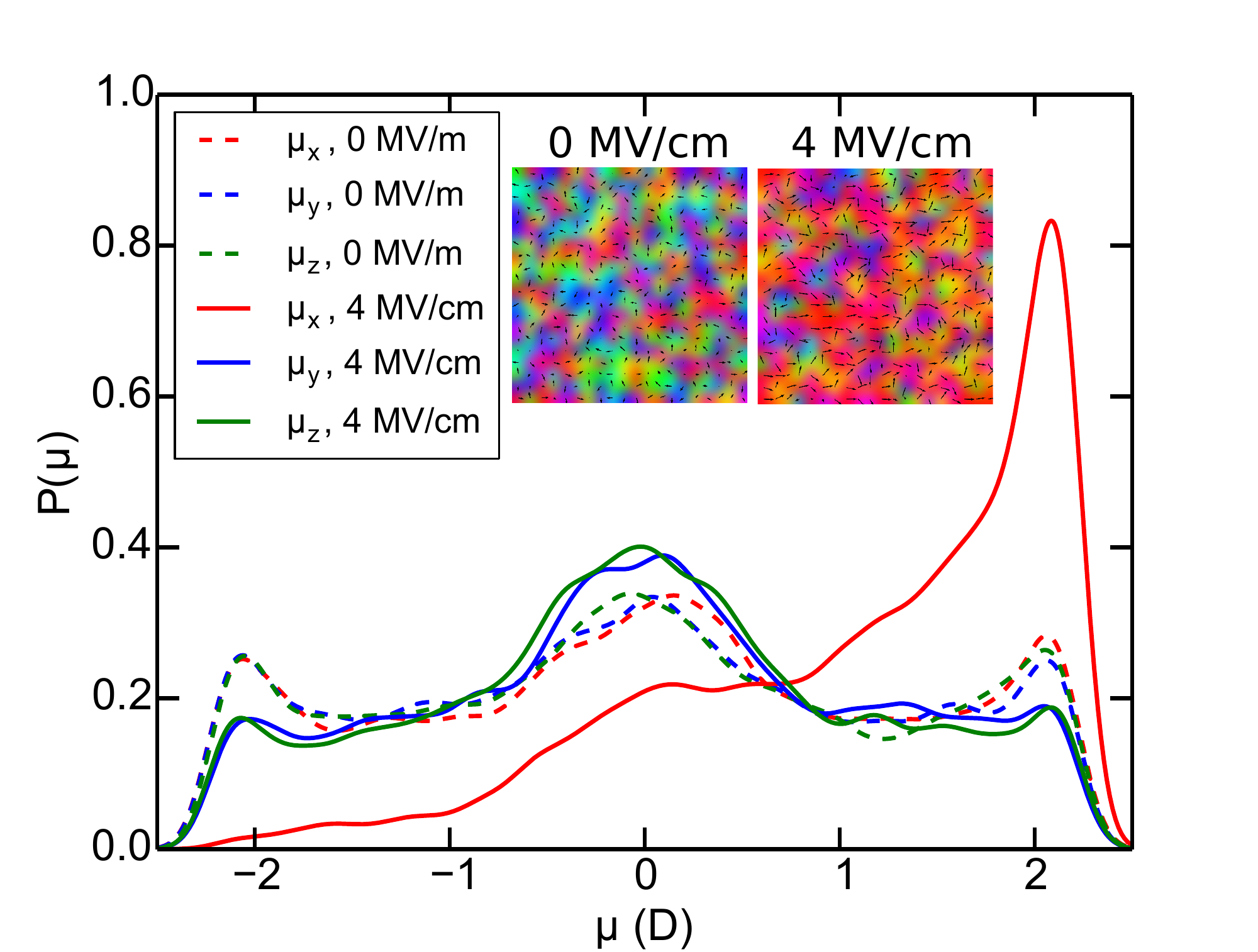}\\
 \caption{Probability distribution functions of molecular dipoles under $E$ = 0 and 4~MV/cm. The inset shows the orientations of molecular dipoles obtained from molecular dynamics simulations at 300~K.}
 \end{figure}
 
\begin{figure}
\centering
\includegraphics[scale=0.5]{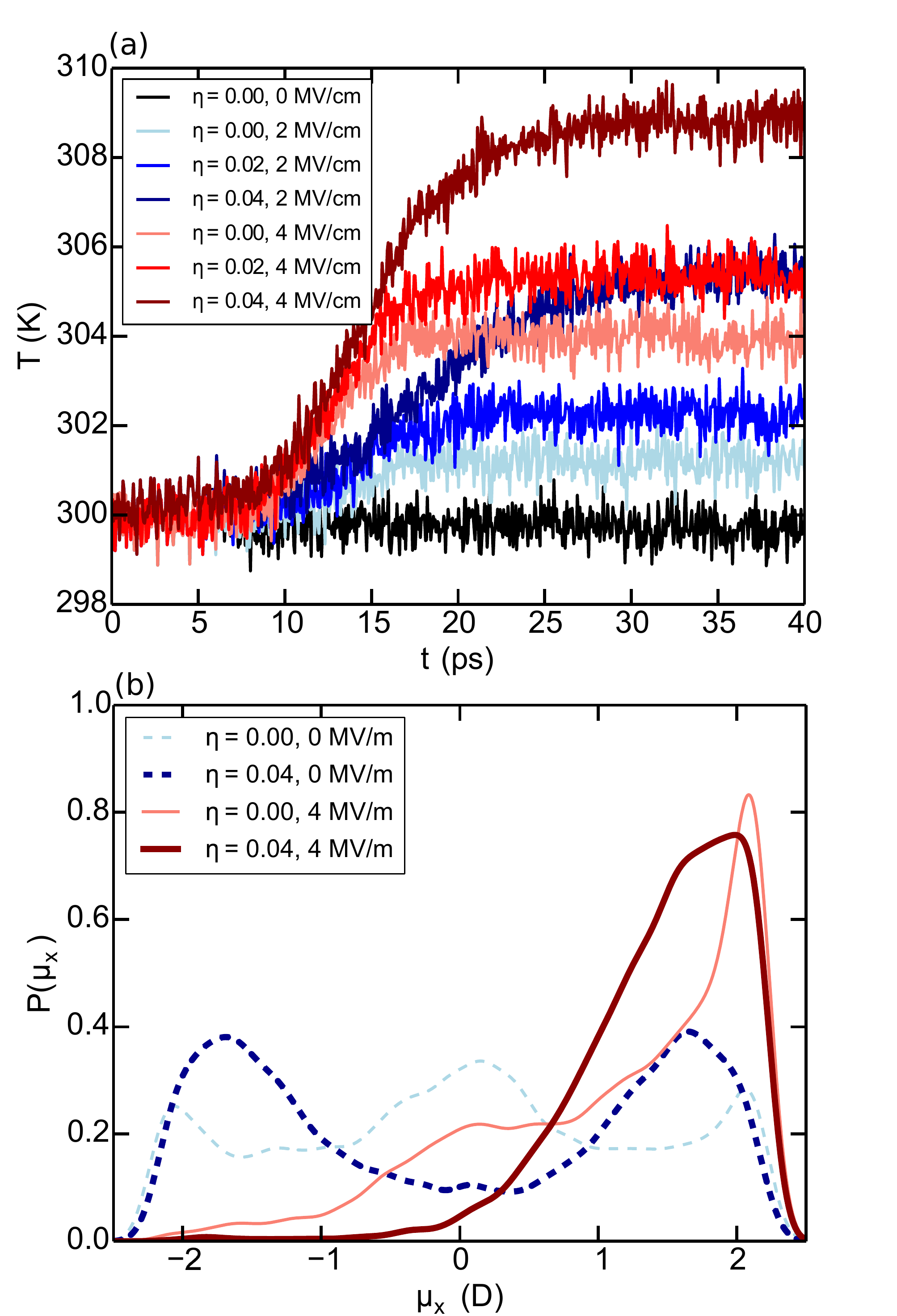}\\
 \caption{Influence of epitaxial strain on (a) electrocaloric effect and (b) probability distribution function of $\mu_x$ in MAPbI$_3$.}
 \end{figure}

\begin{figure}
\centering
\includegraphics[scale=0.5]{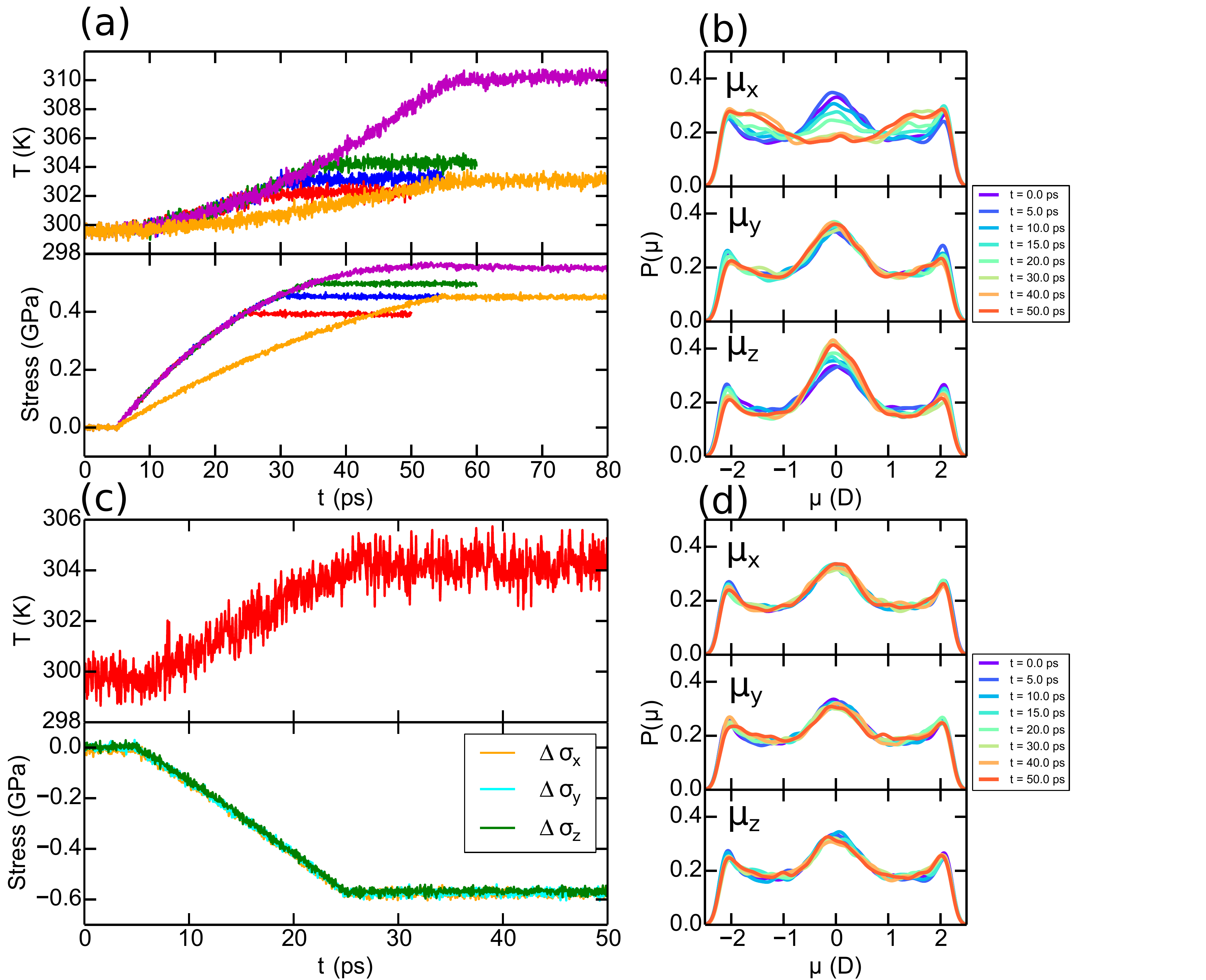}\\
 \caption{Mechanocaloric effect in MAPbI$_3$. (a) Adiabatic thermal change in response to uniaxial stress. (b) Time-resolved probability distribution functions of molecular dipoles for $\Delta \sigma_u=0.45$~GPa. (c) Adiabatic thermal change in response to an isotropic stress. (b) Time-resolved probability distribution functions of molecular dipoles for $\Delta \sigma_i=0.57$~GPa.}
 \end{figure} 
\vspace{-4pt}
\begin{figure}
\centering
\includegraphics[scale=0.4]{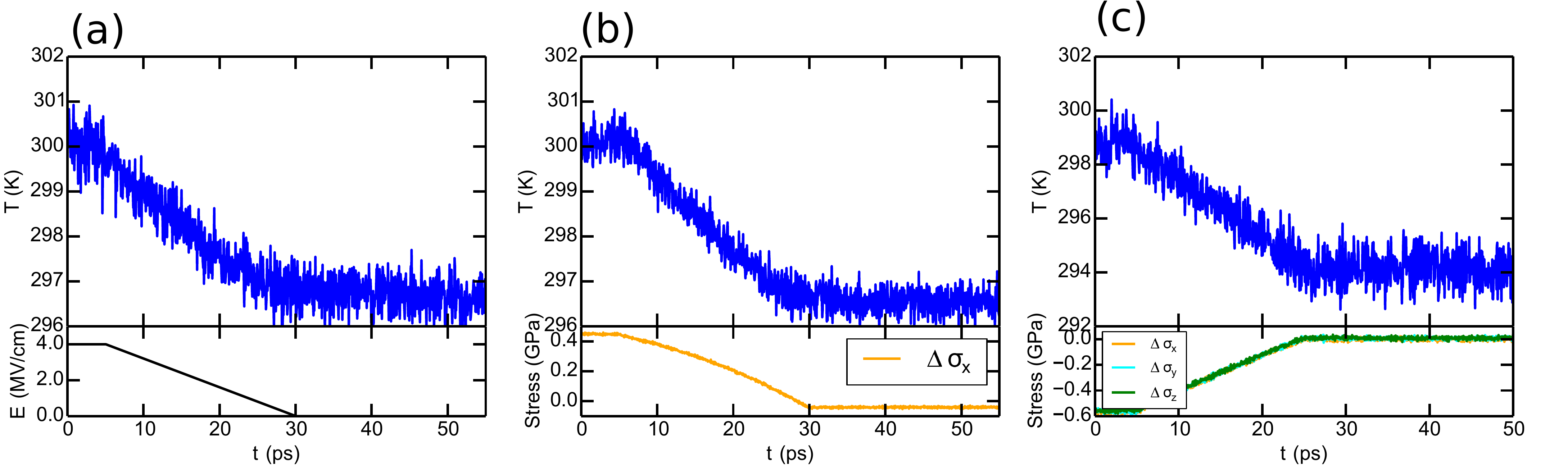}\\
 \caption{Simulated adiabatic cooling during the removal of (a) electric field, (b)uniaxial stress and (c) isotropic stress. }
 \end{figure}  

\end{document}